\documentclass[preprint,aps,nofootinbib]{revtex4-2}
\pdfoutput=1
\usepackage{graphicx}
\usepackage{epstopdf}    
\usepackage{dcolumn}
\usepackage{amsfonts}
\usepackage{amsmath}
\usepackage{amssymb}
\usepackage{bm}
\usepackage{color}
\usepackage{comment}
\usepackage[caption=false]{subfig}
\usepackage{lipsum} 
\usepackage{epsf}
\usepackage{float}
\usepackage{enumitem}
\usepackage[utf8]{inputenc}

\newcolumntype{L}{>{\centering\arraybackslash}m{3cm}}

\newcommand{\edc}{\end{document}}
\newcommand{\bb} {\begin{thebibliography}}
\newcommand{\eb}{\end{thebibliography}}
\newcommand{\bi}[1]{\bibitem{#1}}
\newcommand{\bc}{\begin{center}}
\newcommand{\ec}{\end{center}}

\newcommand{\be}{\begin{equation}\small}
\newcommand{\ee}{\end{equation}\normalsize}
\newcommand{\bea}{\begin{eqnarray}}
\newcommand{\eea}{\end{eqnarray}}
\newcommand{\ba}{\begin{array}{l}   }
\newcommand{\lab}[1]{\label{#1}}
\newcommand{\ea}{\end{array}}

\newcommand{\dsfrac}{\displaystyle\frac}
\newcommand{\ds} {\displaystyle}
\newcommand{\summa}{\ds\sum}
\newcommand{\dssum}{\summa}
\newcommand{\re}[1]{(\ref{#1})}

\newcommand{\ket}[1]{\vert{#1}\rangle}

\newcommand{\ci}{\cite}

\newcommand{\dsint}{\ds\int}

\newcommand{{\vergul}}{  ,}

\newcommand{\veps}{\varepsilon }

\newcommand{\Tcnol}{      {T_{c}^{0}  }                }

\newcommand{\vepsk}{\veps_{k}}

\begin{document}
\draft
\title{Hugenholtz - Pines relations and the critical temperature of a Rabi coupled binary Bose system}

\author{Abdulla Rakhimov$^1$,  Asliddin Khudoyberdiev$^2$ }
	\email{rakhimovabd@yandex.ru, asliddin.khudoyberdiev@tu-dortmund.de}
	\affiliation{$^1$Institute of Nuclear Physics, Tashkent 100214, Uzbekistan \\
	$^2$ Condensed Matter Theory, Department of Physics, TU Dortmund University, 44227 Dortmund, Germany}
\date{\today}

\begin{abstract}
Using a theoretical field Gaussian approximation we have studied
Rabi coupled  binary Bose system at  low temperatures. We have derived
extended Hugenholtz - Pines relations taking into account
one body interaction (e.g. Rabi coupling) and studied the critical temperature $T_c$
of Bose-Einstein condensate transition. We have shown that, the shift of $T_c$ due to this interaction
can not exceed $\sim 60 \%$ and goes to a plateau with increasing the parameter $\Omega_R/T_{c}^{0}$, where
$\Omega_R$ is the intensity of the coupling and $T_{c}^{0}$ is the critical temperature
of the system with $\Omega_R=0$. Moreover, the shift is always positive
and does not depend on the sign of the one body interaction.

\end{abstract}


\maketitle
\section{Introduction}\label{sec1}

The experimental possibility of achieving quantum degeneracy with mixture of atomic gases and binary Bose systems
occupying different hyperfine states or with mixtures of different atomic species has opened
rich opportunities for novel experimental and theoretical studies. The mixtures of atomic gases
 are much more flexible, due to the large variety of available atomic species, characterized by different
hyperfine states, the possibility of generating coherently coupled configurations, and tuning the
interaction between the different components for the mixture \ci{Pitbook14}. For the mixture made of atoms
occupying different hyperfine states, it is possible experimentally to generate coherently coupled
configurations via radio frequency transitions, giving rise to typical Rabi oscillations.
Such experiments were performed with  atoms evaporatively cooled in the
$\ket{F=2, m_F=2}$ and $\ket{F=1,m_F=-1}$ spin states of ${^{87}}$Rb  \ci{hall98,mattews,zibold10}
and very recently, with ${^{39}}$K  atoms in $\ket{F=1, m_F=-1}$ and $\ket{F=1,m_F=0}$ states \ci{lav2021}.
Although, in the most of experimental and theoretical
\ci{lav2021,will00,nied08,xiao13,lelloch13,abad13,aftalion16,salas17,koba19}
studies the existence of binary Bose-Einstein condensate (BEC) has been predicted, possible finite temperature
effects on the properties of the condensates were not considered.

It is well known that a system of bosons, where number of particles is conserved, experiences normal to
BEC phase transition with a certain temperature $T_c$. This critical  temperature essentially depends on internal properties
of the system as well as on the geometry of  the trapping potential. In the simplest case
of a free homogeneous gas of bosonic atoms the critical temperature is given by
$T_{c}^{0}=(2\pi/m)(\rho/\zeta(3/2))^{2/3}$ where $\rho$ is the average particle density
of the Boson gas of atoms with a mass $m$, and $\zeta(n)$ is the Reimann zeta function, $\zeta(3/2)\approx 2.61 $.
The question, attracting for long time attention, is how this expression varies under switching on
one or two body interactions. This problem turned on to be highly non-trivial \ci{yuktc} even
for a one component Bose system. We are motivated by these intesests and analyze binary Bose systems in this work.

The modification of the critical temperature $T_c$ is usually expressed in terms of the relative temperature
shift defined as
\be
\frac{\Delta T_c}{\Tcnol}=\frac{T_c-\Tcnol}{\Tcnol}
\lab{shiftdef}
\ee
whose determination has a long history.
For example, there are several articles, where the authors made an attempt to obtain a proper analytical expression
for the shift due to repulsive contact interaction \ci{baym99,kastening04,ramos01,ouryuk1,ourkleinert,ouriman1,ouriman2},
disorder \ci{ourdisorder,lopatinvinokur}, anisotropic effect in the BEC of triplons \ci{ouraniz1,OurAnn2,ourAniz2part2}
and trap geometry \ci{yuktc}. 
However, in our knowledge, the shift of critical temperature of a binary Bose system
due to the inter-component interaction (with a coupling constant $g_{ab}$
between atoms of $a$ and $b$ components) or especially due to the Rabi coupling has never been studied.

The goals of  the present work are derivation
Hugenholtz - Pines (HP) relations and
  determination of the behavior of the transition temperature
of  binary Bose system in the presence of a one body interaction implemented by optical Rabi coupling.
For this purpose we use Gaussian (one loop ) approximation \ci{salas17,hague,andersen} with Rabi coupling in the framework of the mean field
theory and derive  an analytical expression for the Rabi induced shift of the critical temperature. We will show that
the shift is positive and increases toward an asymptotic value with increasing the strength of Rabi coupling.
During the calculations we shall derive also Hugenholtz -Pines relations for this system, which is essential for the BEC with a gapless spectrum.

This work is organized as follows. In Sec. II starting from the hamiltonian with one
and two body interactions, we derive explicit expressions for the Green functions and self energies
 in order to find HP relations in Gaussian approximation. In Sec. III we discuss extremums of the
free energy and study conditions for the existence of a pure BEC in an equilibrium state of
a Rabi coupled binary Bose systems. In Sec. IV we concentrate on a symmetric binary Bose system
to study BEC and normal phases separately. The critical temperature and its shift
will be studied in Sec. V , where we present our numerical results also. In the last section we summarize
our findings.

\section{Hugenholtz-Pines relations}

We start with the grand canonical Hamiltonian for homogenous binary Bose system with Rabi coupling
\ci{lelloch13}:
\begin{subequations}
\begin{align}
&\hat{H}=\hat{H_1}+\hat{H_2}+\hat{H}_{12}  \label{Htot}\\
& \hat{H_a}=\dsint d^3r
\left[
-\psi^{\dagger}_{a}  \frac {{{\nabla}}^2} {2m}\psi_{a}-\mu_a \psi^{\dagger}_{a}\psi_{a}+
\frac{g_a}{2}(\psi^{\dagger}_{a}\psi_{a})^2
         \right]\label{Ha} \\
 &\hat{H}_{12}=\dsint d^3r
\left[
g_{12}\psi^{\dagger}_{1}\psi^{\dagger}_{2}\psi_1\psi_2+
 \frac {\Omega_R}{2}(\omega_R\psi^{\dagger}_{1}\psi_{2}+h.c.) \right]
%
\label{Hab}
\end{align}
\end{subequations}
where  the associated chemical potentials are represented by $\mu_{a}$,  while $m$
represents  the atomic  mass. Due to the flipping term only the total number of particles
(total density in the uniform system)  is
conserved. Thus the chemical potential should be  the same for
both components: $\mu_1=\mu_2=\mu$. In terms of the corresponding $s$-wave scattering lengths $a_s$,
the coupling constants can be written as $g_{a}=4\pi a_{a}/m$, while the cross
coupling is $g_{12}=4\pi a_{12}/m$, and $\psi_a(\textbf{r})$, ($a=1,2$),  is the bosonic field
operator of the component $a$ of the spinor $\psi=(\psi_1,\psi_2)^{T}$  .
 Here and below we set $\hbar=1$, $k_B=1$.

The coherent (Rabi) coupling is given by  the last term of \re{Hab} with the intensity $\Omega_R\geq 0$
and the phase $\omega_R=\exp{(i\theta_R)}$.
Depending on the physical system,
this term can have its origin  on  a two-photon
(Raman) process also.
It is clear that, when $\Omega_R=0$ , the phase of each component is  independent and
${\hat H}$ is invariant under the transformations $\{\psi_{1}\rightarrow \exp(i\theta_1)\psi_{1}$,
$\psi_{2}\rightarrow \exp(i\theta_2)\psi_{2}\}$.
 The spontaneous breaking of this invariance leads
to the emergence of Goldstone (gapless) modes  for both components. However, when $\Omega_R\neq 0$
the Hamiltonian \re{Htot} is invariant under the unique gauge  transformation
 $\{\psi_1\rightarrow \exp(i\theta_0)\psi_1 ,
\psi_2\rightarrow \exp(i\theta_0)\psi_2 \}$ with the same  phase angle $\theta_1=\theta_2=\theta_0$.
In this case the spontaneous breaking of this symmetry makes at least one of  branches of excitation spectrum
as gapless.

In fact, the particle spectrum under the spontaneously  broken gauge symmetry has to be gapless.
This is , actually, one of the main conditions for the existence of a stable BEC .
Since if there would be a gap in the spectrum, there could be no macroscopic occupation
of a single ground state level.  Hugenholtz and Pines \ci{hp}, (and later Bogolyubov \ci{bogpines})
showed that for one component Bose system the chemical potential is expressed through the normal
$\Sigma_n({\mathbf{k}},\omega)$
and anomalous $\Sigma_{an}({ \mathbf{k}},\omega)$
 self energies as
$\mu=\Sigma_n(0,0)-\Sigma_{an}(0,0)$. Further, this relation has been extended \ci{nepom,watabe21} for
a two component Bose system with a two body coupling.
The question arises, how the presence of a one body coupling, (e.g. Rabi coupling) in a binary Bose system
will modify   Hugenholtz-Pines  relations?

In the next section we make an attempt to find an answer to this question, at
least on the level of the Gaussian (bilinear) approximation \ci{hague,stoof},
which is a particular case of a more accurate approach as Hartree - Fock - Bogoliubov
\ci{ourtan1,ourtan2}.
For this purpose we  derive explicit expressions for Green functions,   as well as  for the
particle spectrum and find relations between the  self
energies. Note that, particularly, for  a one component Bose system
HP relation can be directly obtained by the condition of the existence of the Goldstone mode.
\subsection{The Green functions}
The finite temperature Euclidean $(\tau=it)$ space time action, corresponding to the Hamiltonian
\re{Htot} is given by
\begin{eqnarray} \nonumber
&S=\int_0^\beta d\tau\int d^3r[\psi_{1}^\dagger\hat{K}\psi_{1}
+\psi^{\dagger}_{2}\hat{K}\psi_{2}+\frac{g_1}{2}(\psi^{\dagger}_{1}\psi^{\dagger}_{1}\psi_{1}\psi_{1})+ \\ \nonumber
&\frac{g_2}{2}(\psi^{\dagger}_{2}\psi^{\dagger}_{2}\psi_{2}\psi_{2})+
g_{12}(\psi^{\dagger}_{1}\psi_{1})
(\psi^{\dagger}_{2}\psi_{2})+ \\
&\frac{\Omega_R}{2}(\omega_R\psi^{\dagger}_{1}\psi_{2}+\omega_R^\star\psi^{\dagger}_{2}\psi_{1})]
\label{action}
\end{eqnarray}
where $\hat {K}=\partial/\partial \tau-{\bf{\nabla}}^2/2m-\mu$ .
In Eq. (\ref{action}) the fields $\psi_1(\mathbf{r},\tau)$ and $\psi_2(\mathbf{r},\tau)$ are
periodic in $\tau$ with the period $\beta=1/T$.
Now  we introduce Bogoliubov shift
\bea
\psi_{1}=\sqrt{\rho_{0a}}+\tilde{\psi}_{1}, \quad   \psi_{2}=\xi\sqrt{\rho_{0b}}+\tilde{\psi}_{2}
\label{shift}
\eea
where $\rho_{0a,b}$ is the condensate fraction of the component $a$, $(b)$, $\xi=e^{i\theta}$ with the relative phase angle $\theta$ between two Bose-Einstein condensates and $\tilde{\psi_{1}}$ and
$\tilde{\psi}_{2}$ are complex fluctuating fields which will be integrated out
\footnote{By default $a$ and $b$ correspond to the first and the second components
of the spinor, respectively. For example, $g_a\equiv g_1$,$g_b\equiv g_2$ and $g_{ab}\equiv g_{12}$}.
Due to the \re{Hab} term in the Hamiltonian, both components are coherently coupled, and hence
the relative phase $\xi$ should be determined from the minimum condition of the thermodynamic potential.
On the other hand, it has been proven \ci{OurAnn2,ourAniz2part2} that  the phase angle of a pure BEC should be
equal to $\pi n$ with an integer $n$. Consequently, the relative phase should be real, $\xi=\pm 1$.
It can be shown that
 for a homogenous binary Bose system a
physical observable should not depend on the relative phase $\xi$, or more precisely on its sign:
$\xi=+1$  or $\xi=-1$,
\footnote{This rule may be referred as a phase invariance.} by the condition
$\xi\omega_R=-1$ (see Appendix). So, for the convenience we set $\xi=1$, $\omega_R=-1$ in the
rest of the paper.

 Note that the Bogoliubov shift is an
exact canonical transformation \cite{Yukalov_LP},
and
not an approximation, as sometimes
it is stated. For a uniform system at equilibrium, $\rho_{0a}$ and $\rho_{0b}$ are real
variational constants, which are fixed by the minimum of the free energy $\Omega$ as
$\partial\Omega/\partial\rho_{0,a,b}=0$,
$\partial^2\Omega/\partial^2\rho_{0,a,b}\geqslant 0$ \cite{OurAnn2}.
As to the numbers of uncondensed particles $N_{1a}$ and $N_{1b}$, they are related to
the fields $\tilde{\psi}_1$ and $\tilde{\psi}_2$:
\bea
&N_{1a}=V\rho_{1a}=\int d^3r\langle\tilde{\psi}^{\dagger}_{1}(\mathbf{r})\tilde{\psi}_{1}(\mathbf{r})\rangle \\ \nonumber
\\ 
&N_{1b}=V\rho_{1b}=\int d^3r\langle\tilde{\psi}^{\dagger}_{2}(\mathbf{r})\tilde{\psi}_{2}(\mathbf{r})\rangle \; 
\label{n1a1b}
\eea
so that
\bea
&&N_a=\int d^3r\langle \psi^{\dagger}_{1}(\mathbf{r})\psi_{1}(\mathbf{r}) \rangle \\
&&N_b=\int d^3r\langle \psi^{\dagger}_{2}(\mathbf{r})\psi_{2}(\mathbf{r})
\label{n1a1b1}
\rangle
\eea
with the normalization conditions $N=N_a+N_b$, $N_a=V\rho_a=V(\rho_{0a}+\rho_{1a})$,
and $N_b=V\rho_b=V(\rho_{0b}+\rho_{1b})$, where $N_{a(b)}$ is the number of particles in the
component $a$, ($b$), $N$ is the particle number in the whole two-component system confined in the volume
$V$ .
Since we are considering a homogeneous system, the densities $\rho_a$ and $\rho_b$ are
uniform.

The mean-field plus Gaussian approximation is obtained by expanding $S$ up to the second order in fluctuating fields \ci{salas17}. So,  inserting (\ref{shift}) into (\ref{action}), we represent the effective action as



\begin{subequations}
\begin{align}
&S\approx S_0+S_2  \label{Stot}\\
&S_0=\int_0^\beta d\tau\int d^3r\{\hat{K}\rho_{0a}+\hat{K}\rho_{0b}+\frac{g_a}{2}\rho_{0a}^2+
\frac{g_b}{2}\rho_{0b}^2+g_{ab}\rho_{0a}\rho_{0b}-
{\Omega_R\sqrt{\rho_{0a}\rho_{0b}}}\} \label{S0}\\
&S_2=-\frac{1}{2}\int d\tau d\tau'd^3rd^3r'[\tilde{\psi}^{\dagger}_{1}(\tau,\textbf{r}),
\tilde{\psi}_{1}(\tau,\textbf{r}),\tilde{\psi}^{\dagger}_{2}
(\tau,\textbf{r}),\tilde{\psi}_{2}(\tau,\textbf{r})]
D^{-1}(\tau,\textbf{r},\tau',\textbf{r}')
\begin{vmatrix}
     \tilde{\psi}_{1}(\tau',\textbf{r}')\\
     \tilde{\psi}^{\dagger}_{1}(\tau',\textbf{r}')\\
     \tilde{\psi}_{2}(\tau',\textbf{r}')\\
     \tilde{\psi}_{2}^\dagger(\tau',\textbf{r}')
\end{vmatrix}
\end{align}
\end{subequations}
where the inverse Green function in momentum space is given by
\bea \label{mat1}
&D^{-1}(\mathbf{k},\omega_n)= \\ \nonumber
&\begin{pmatrix}
     i\omega_n-\varepsilon_k+\Lambda_a& -g_a\rho_{0a}& -g_{ab}\sqrt{\rho_{0a}\rho_{0b}}+\frac{\Omega_R}{2}&
 -g_{ab}\sqrt{\rho_{0a}\rho_{0b}}\\
     -g_a\rho_{0a}& -i\omega_n-\varepsilon_k+\Lambda_a& -g_{ab}\sqrt{\rho_{0a}\rho_{0b}}&
-g_{ab}\sqrt{\rho_{0a}\rho_{0b}}+\frac{\Omega_R}{2}\\
      -g_{ab}\sqrt{\rho_{0a}\rho_{0b}}+\frac{\Omega_R}{2}& -g_{ab}\sqrt{\rho_{0a}\rho_{0b}}&
 i\omega_n-\varepsilon_k+\Lambda_b & -g_b\rho_{0b}\\
      -g_{ab}\sqrt{\rho_{0a}\rho_{0b}}& -g_{ab}\sqrt{\rho_{0a}\rho_{0b}}+\frac{\Omega_R}{2}& -g_b\rho_{0b} & -i\omega_n-\varepsilon_k+\Lambda_b
&\end{pmatrix}
\eea

where $\varepsilon_k={ k}^2/2m$, $\Lambda_a=\mu-2g_a\rho_{0a}-g_{ab}\rho_{0b}$; $\Lambda_b=\mu-2g_b\rho_{0b}-g_{ab}\rho_{0a}$.
Now we define self-energies as \ci{stoof}
\bea \label{self}
{\Sigma}_{ij}=(D^{-1}_0)_{ij}-D^{-1}_{ij}
\eea
where ${\hat D}_0$ corresponds to the "ideal gas" with Rabi coupling:

\bea \label{mat2}\nonumber
&D^{-1}_0(\mathbf{k},\omega_n)=D^{-1}(\mathbf{k},\omega_n)\vert_{g_{a,b}=0,g_{ab}=0}=\\
&\begin{pmatrix}
     i\omega_n-\varepsilon_k+\mu& 0 & \frac{\Omega_R}{2}& 0\\
     0& -i\omega_n-\varepsilon_k+\mu& 0& \frac{\Omega_R}{2}\\
      \frac{\Omega_R}{2}& 0& i\omega_n-\varepsilon_k+\mu & 0\\
      0& \frac{\Omega_R}{2}& 0 & -i\omega_n-\varepsilon_k+\mu
&\end{pmatrix}
\eea

Note that,  the  Green function $\hat{D}_0$ may be used  in organizing perturbative scheme in terms of coupling constants. From \re{mat1}, (\ref{self}) and (\ref{mat2}) we immediately  obtain
\bea \label{mat3}
\hat{\Sigma}(\mathbf{k},\omega_n)=
\begin{pmatrix}
     2g_a\rho_{0a}+g_{ab}\rho_{0b}& g_a\rho_{0a} & g_{ab}\sqrt{\rho_{0a}\rho_{0b}}& g_{ab}\sqrt{\rho_{0a}\rho_{0b}}\\
     g_a\rho_{0a}& 2g_a\rho_{0a}+g_{ab}\rho_{0b}&  g_{ab}\sqrt{\rho_{0a}\rho_{0b}}& g_{ab}\sqrt{\rho_{0a}\rho_{0b}}\\
       g_{ab}\sqrt{\rho_{0a}\rho_{0b}}&  g_{ab}\sqrt{\rho_{0a}\rho_{0b}}& 2g_b\rho_{0b}+g_{ab}\rho_{0a}& g_b\rho_{0b}\\
       g_{ab}\sqrt{\rho_{0a}\rho_{0b}}& g_{ab}\sqrt{\rho_{0a}\rho_{0b}}& g_b\rho_{0b} & 2g_b\rho_{0b}+g_{ab}\rho_{0a}
\end{pmatrix}
\eea
Thus normal and anomalous self-energies are
\bea \nonumber
&\Sigma_n^{a}=2g_a\rho_{0a}+g_{ab}\rho_{0b} \quad \\
&\Sigma_n^{b}=2g_b\rho_{0b}+g_{ab}\rho_{0a} \\
&\Sigma_{an}^{a}=g_a\rho_{0a}, \quad  \Sigma_{an}^{b}=g_b\rho_{0b}\\
&\Sigma_n^{ab}= \Sigma_{an}^{ab}=g_{ab}\sqrt{\rho_{0a}\rho_{0b}}
\label{selfs}
\eea
On the other hand, the Green functions and excitations spectrum can be presented in a more compact form in Cartesian representation
(real field formalism) as
\bea \label{cartes}
\tilde{\psi}_1=\frac{1}{\sqrt{2}}(\phi_1+i\phi_2),
\quad \tilde{\psi}_2=\frac{1}{\sqrt{2}}(\phi_3+i\phi_4)
\eea
with real functions $[\phi_1,\phi_2, \phi_3, \phi_4]$, such that
\be
\ba
S_2=  
\dsfrac{1}{2}\dsint d\tau d^3rd\tau' d^3r'\dssum_{i,j=1}^4
\phi_i(\tau,\textbf{r})G_{ij}^{-1}(\tau, \textbf{r},\tau', \textbf{r}')\phi_j(\tau',\textbf{r}')
\label{act2}
\ea
\ee

Now from (\ref{action}), \re{selfs}, (\ref{cartes}) and (\ref{act2}) one finds
\bea \label{mat4}
&G^{-1}(\mathbf{k},\omega_n)=\\\nonumber
&\begin{pmatrix}
    \varepsilon_k+X_1 & \omega_n& X_5 & 0\\
     -\omega_n & \varepsilon_k+X_2 & 0 & X_6\\
      X_5 & 0 & \varepsilon_k+X_3 & \omega_n\\
      0 & X_6 & -\omega_n & \varepsilon_k+X_4
\end{pmatrix}
\eea
where
\bea \label{xlar}\nonumber
X_1=&-\mu+3\rho_{0a}g_a+\rho_{0b}g_{ab}=\Sigma_n^a+\Sigma_{an}^a-\mu \\ \nonumber
X_2=&-\mu+\rho_{0a}g_a+\rho_{0b}g_{ab}=\Sigma_n^a-\Sigma_{an}^a-\mu \\  \nonumber
X_3=&-\mu+3\rho_{0b}g_b+\rho_{0a}g_{ab}=\Sigma_n^b+\Sigma_{an}^b-\mu \\ \nonumber
X_4=&-\mu+\rho_{0b}g_b+\rho_{0a}g_{ab}=\Sigma_n^b-\Sigma_{an}^b-\mu \\ 
X_5=&2 g_{ab}\sqrt{\rho_{0a}\rho_{0b}}-\frac{\Omega_R}{2}, \quad 
X_6=-\dsfrac{\Omega_R}{2}  \quad
\eea
 so that $\Sigma_n^{ab}=\Sigma_{an}^{ab}=g_{ab}\sqrt{\rho_{0a}\rho_{0b}}=(X_5-X_6)/2$.

From the condition $\det [G^{-1}(\mathbf{k},\omega_n)]=0$ one may find the excitation spectrum:
\be \lab{om12gen}
\ba
\omega_1^2=\dsfrac{E_a^2+E_b^2}{2}+X_5X_6+\dsfrac{\sqrt{D_s}}{2} \\
\omega_2^2=\dsfrac{E_a^2+E_b^2}{2}+X_5X_6-\dsfrac{\sqrt{D_s}}{2}
\ea
\ee
where
\be
\ba
D_s=(E_a^2-E_b^2)^2+4X_5X_6(E_a^2+E_b^2)
+4X_6^2E_{13}^2+4X_5^2E_{24}^2,\\
E_a^2=W_1W_2,\quad  E_b^2=W_3W_4, \quad E_{13}^2=W_1W_3, \\
E_{24}^2=W_2W_4  , \quad W_i=\varepsilon_k+X_i, \quad (i=1,2,3, 4)
\label{Ds}
\ea
\ee

Now we come back to the question on HP relations. For simplicity we limit ourselves
to the symmetrical case with $g_a=g_b=g$, $\rho_{0a}=\rho_{0b}=\rho_0/2$ ,
$\Sigma_n^a=\Sigma_n^b=\Sigma_n$ and $\Sigma_{an}^a=\Sigma_{an}^b=\Sigma_{an}$,
to rewrite the dispersions in \re{om12gen} as
\be
\ba
\omega_1^2=(\varepsilon_k+X_2+X_6)(\varepsilon_k+X_1+X_5) \\
\omega_2^2=(\varepsilon_k+X_2-X_6)(\varepsilon_k+X_1-X_5)
\lab{om12sym1}
\ea
\ee
Both from experimental and theoretical studies it is well known \ci{Pitbook14}
that, spontaneous symmetry breaking of gauge invariance in the Hamiltonian
\re{Htot} leads to the appearance of two modes, $\omega_d$ and $\omega_s$ ,
at that the density mode $\omega_d$ is gapless, while the spin mode $\omega_s$ has a finite gap.
From Eq. \re{om12sym1} it is seen that this may be achieved by imposing a simple condition as $X_2=X_6$,
which in terms of the self energies will be equivalent to the relation

\be
\mu=\Sigma_n-\Sigma_{an}-\frac{\Omega_R}{2}
\lab{hp1}
\ee

where we used Eq. \re{xlar}. Note that, in the absence of the Rabi coupling one comes back to the
familiar HP relation $\Sigma_n-\Sigma_{an}=\mu$. In the next section we show
that the chemical potential determined by the relation \re{hp1} indeed will
correspond to the minimum of the  thermodynamic potential in an  equilibrium.

\section{Thermodynamic potential and the condition for the existence of a pure BEC}
The grand canonical potential can be easily found as
\bea \label{pot} \nonumber
\Omega=-T\ln Z, \qquad 
 Z=\dsint D\phi_1D\phi_2D\phi_3D\phi_4e^{-S[\phi_1,\phi_2,\phi_3,\phi_4]}
\eea
where $S=S_0+S_2$  with $S_0$ and $S_2$ are given in Eqs. \re{S0} and \re{act2}, respectively.
The path integral in \re{pot} is Gaussian and can be evaluated exactly.
As a result one obtains:
\bea \label{omegat}
\Omega=\Omega_0+\Omega_{ln}
\eea
\bea \label{omega0} \nonumber
\Omega_0=V[-\mu\rho_{0a}-\mu\rho_{0b}+\frac{g_a\rho_{0a}^2}{2}+\frac{g_b\rho_{0b}^2}{2}+
g_{ab}\rho_{0a}\rho_{0b}-{\Omega_R\sqrt{\rho_{0a}\rho_{0b}}}]
\lab{om0}
\eea
\be
\ba
\Omega_{ln}=\dsfrac{1}{2}\dssum_k[\omega_1+\omega_2-2\varepsilon_k+
(counter \quad terms)]
+T\dssum_k\ln(1-e^{-\beta\omega_1})+T\dssum_k\ln(1-e^{-\beta\omega_2})
\label{omegaln}
\ea
\ee
where $\omega_{1,2}$ are given in Eqs. (\ref{om12gen}) and (\ref{Ds}).
Therefore,  saddle - point  equations  (\ref{equlib}) \ci{salas17,hague,andersen} will  have following  explicit form:
\be \label{eqrez}
 \begin{split}
\dsfrac{\partial\Omega_0}{\partial\rho_{0a}}=-{V}[\mu-g_a\rho_{0a}-g_{ab}\rho_{0b}+
\dsfrac{\Omega_R\rho_{0b}}{2\sqrt{\rho_{0a}\rho_{0b}}}]=0 \\
\dsfrac{\partial\Omega_0}{\partial\rho_{0b}}=-{V}[\mu-g_b\rho_{0b}-g_{ab}\rho_{0a}+\dsfrac{\Omega_R\rho_{0a}}{2\sqrt{\rho_{0a}\rho_{0b}}}]=0
\end{split}
\ee
which give in particular
\be
\mu=g_a\rho_{0a}+g_{ab}\rho_{0b}-\dsfrac{\Omega_R\rho_{0b}}{2\sqrt{\rho_{0a}\rho_{0b}}}
\label{eq:mu}
\ee
From this equation one may come to the following important conclusion:
 If $\rho_{0a}(T)\neq\rho_{0b}(T)$, then there would be no pure  BEC
with a certain critical temperature above which $\rho_{0a}(T>T_c)$ or $\rho_{0b}(T>T_c)$
absolutely vanish.
Instead, since the chemical potential should be  finite, $(\mu\neq\infty)$, one has to deal with
 a crossover from BEC to the normal phase transition.
The situation may be cured by assumption $\rho_{0a}(T)=\rho_{0b}(T)$ in the whole ranges of
the temperatures. Clearly this can take place in the symmetric case with equal masses and coupling constants.
For this reason in the rest of  this work we shall consider only symmetric binary Bose system.

\section{Symmetric binary Bose systems with one body coupling}

Let $\rho_a=\rho_b=\rho/2$, $\rho_{0a}=\rho_{0b}=\rho_0/2$, $g_a=g_b=g$,
 and
hence $X_3=X_1$, $X_4=X_2$.
Below we first derive explicit expressions for the density of the uncondensed fraction
$\rho_{1a}=\rho_{1b}=\rho_{1}/2$ and then discuss its features in the normal
and BEC phases.
\subsection{Uncondensed fraction $\rho_1=\rho-\rho_0$}.

It is well known that quantum and temperature fluctuations tend to destroy BEC
leading to an emergence of a depletion  defined as
\be
\ba
\rho_{1a}=\dsfrac{1}{V}\dsint d^3r\langle \Tilde{\psi}^{\dagger}_{a}
({\bf{r}})\Tilde{\psi}_a({\bf{r}})\rangle=
\dsfrac{1}{2V}\dsint d^3r[G_{11}(\mathbf{r},\tau;\mathbf{r}^\prime, \tau^\prime)+ 
G_{22}(\bf{r},\tau;\mathbf{r}^\prime, \tau^\prime)]
\Big\vert_{\mathbf{r}-\mathbf{r}^\prime\rightarrow 0,\tau-\tau^\prime\rightarrow 0}
\label{rho1a1}
\ea
\ee
 where \footnote{For the component $b$ we have the similar expression}
\be
\ba
G_{ij}(\mathbf{r},\tau;\mathbf{r}^\prime, \tau^\prime)= 
\frac{1}{V\beta}\sum_{\omega_n,\mathbf{k}}e^{i\mathbf{k}(\mathbf{r}-\mathbf{r}^\prime)}e^{i\omega_n(\tau-\tau^\prime)}G_{ij}(\omega_n,\mathbf{k})  
\lab{grfur}
\ea
\ee
and $\omega_n=2\pi n T$,  is the Matsubara frequency.
The matrix elements of the Green function $G_{ij} (x,x) $ can be obtained by inverting
${\hat G}^{-1}$ in Eq. \re{mat4} and performing Matsubara summations in \re{grfur}. The result is given by
\be
\ba
G_{11}(x,x)=G_{33}(x,x)=A_1(\mathbf{k},\omega_1,X_1,X_2,X_5,X_6)+
A_1(\mathbf{k},\omega_2,X_1,X_2,-X_5,-X_6)\\
G_{22}(x,x)=G_{44}(x,x)=A_2(\mathbf{k},\omega_1,X_1,X_2,X_5,X_6)+
A_2(\mathbf{k},\omega_2,X_1,X_2,-X_5,-X_6)\\
G_{13}=A_1(\mathbf{k},\omega_1,X_1,X_2,X_5,X_6)- 
A_1(\mathbf{k},\omega_2,X_1,X_2,-X_5,-X_6)\\
G_{24}=A_2(\mathbf{k},\omega_1,X_1,X_2,X_5,X_6)- 
A_2(\mathbf{k},\omega_2,X_1,X_2,-X_5,-X_6)
\lab{gxxsym}
\ea
\ee
 where $x\equiv (\mathbf{r},\tau)$ and the rest of matrix elements equals to zero, e.g. $G_{12}(x,x)=0$.
Here we have introduced following notations
\be
\ba
A_1(\mathbf{k},\omega_1,X_1,X_2,X_5,X_6)=
\dssum_{\mathbf{k}}\left[\dsfrac{X_{6}^{2}W_1+X_{5}W_{2}^{2}+W_2X_5X_6+W_1W_2X_6
}{2\omega_1(W_2X_5+W_1X_6)}W(\omega_1)\right]\\
A_2(\mathbf{k},\omega_1,X_1,X_2,X_5,X_6)=
\dssum_{\mathbf{k}}\left[\dsfrac{X_{5}^{2}W_2+X_{6}W_{1}^{2}+W_1X_5X_6+W_1W_2X_5
}{2\omega_1(W_2X_5+W_1X_6)}W(\omega_1)\right]
\lab{Alar}
\ea
\ee
where  $W(\omega)=1/2+1/(\exp{(\omega\beta})-1)$ and $\omega_{1,2}$ are
defined in the Eqs. \re{om12sym1}.
Now from Eqs. \re{rho1a1}-\re{Alar} one can immediately find
\be
\ba
\rho_{1a}=\dsfrac{1}{2V}\dssum_{\mathbf{k}}
\left[
\dsfrac{X_{6}^{2}W_1+X_{5}^2W_2}{2\omega_1(W_2X_5+W_1X_6)}W(\omega_1)\right]+
\dsfrac{1}{2V}\dssum_{\mathbf{k}}
\left[
\dsfrac{(W_1+W_2)(W_1X_6+W_2X_5+X_5X_6)
}{2\omega_1(W_2X_5+W_1X_6)}W(\omega_1)\right]+\\
\dsfrac{1}{2V}\dssum_{\mathbf{k}}\left[(\omega_1\rightarrow \omega_2,X_5\rightarrow -X_5,X_6\rightarrow -X_6)-1\right]
=\rho_{1b}
\lab{rho1aany}
\ea
\ee
where $X_1 \dots X_6$ and the dispersions $\omega_{1,2}$ are given in Eq.s \re{Xn} and \re{dispany}.
 Below we discuss BEC and normal states separately.

\subsection{Normal phase ($T>T_c$).}
Here we have $\rho_{0a}=\rho_{0b}=0$ and Eqs. \re{Xn} give

\be
\ba
X_1(T>T_c)=X_2(T>T_c)=-\mu \\
 X_5(T>T_c)=X_6(T>T_c)=-\Omega_R/2
\lab{Xlartbig}
\ea
\ee
and hence
\be
\ba
\omega_{1}^2(T>Tc)=(\veps_k+X_1+X_5)^2=(\varepsilon_k-\mu-\Omega_R/2)^2\\
\omega_{2}^2(T>Tc)=(\veps_k+X_1-X_5)^2=(\varepsilon_k-\mu+\Omega_R/2)^2
\lab{disptbig}
\ea
\ee
The density of uncondensed particles, say of the component $a$, given by the general expression
\re{rho1aany} will be exactly equal to the density of the particles of the sort $a$
with the following explicit expression
\be
\rho_{1a}(T>T_c)=\rho_a=\frac{\rho}{2}
=\frac{1}{2V}\sum_k\left[\frac{1}{e^{\beta\omega_1}-1}+\frac{1}{e^{\beta\omega_2}-1}\right]
\lab{rho1atbig}
\ee
where $\omega_{1,2}$ are defined in Eqs. \re{disptbig}.
Note that in the normal state HP relations  do not make sense, so the chemical
potential $\mu(T>T_c)$ should be evaluated as the solution to the equation
\re{rho1atbig} with the given input parameters such as $\rho$, $T$ and $\Omega_{R}$.

\subsection{Condensed phase.}
On the other hand, at enough low temperatures, $T<T_c$, HP relations are appropriate.
Moreover, it can be easily shown that, the chemical potential $\mu$ determined by
the HP relation in  Eq. \re{hp1} corresponds to the minimum of $\Omega$.
In fact, $\mu$ in Eq. \re{eq:mu} can be rewritten as
\be
\mu=g\rho_{0a}+g_{ab}\rho_{0a}-\frac{\Omega_R}{2}
\lab{musym}
\ee
in the symmetric case. Now inverting the  explicit expressions for the self energies
\re{selfs}, one finds
\be
g\rho_{0a}=\Sigma_{n}, \quad \quad g_{ab}\rho_{0a}=\Sigma_{n}-2\Sigma_{an}
\ee
and inserting these into  \re{musym} comes back  exactly to   the HP relations in  Eq.   \re{hp1}.
Thus we conclude that HP relations for a symmetric binary Bose system with Rabi coupling, introduced by the hamiltonian
\re{Hab} are presented as
\be
\ba
\Sigma_{n}-\Sigma_{an}=\mu+\frac{\Omega_R}{2}, \quad\quad \Sigma_{n}^{ab}=\Sigma_{an}^{ab}
\lab{hpfinal}
\ea
\ee

As it has been pointed out in Sect.II, due to these relations the excitation spectrum has
two branches:
one is gapless with
\be 
\ba
\omega_{1}^{2}=\vepsk(\vepsk+X_1+X_5)= 
\vepsk[\vepsk+2\rho_{0a}(g+g_{ab})]\approx ck^2+O(k^4)
\lab{om1bec}
\ea
\ee
and the other one with a gap:
\be
\omega_{2}^{2}=[\vepsk+\Omega_R][\vepsk+\Omega_R+2\rho_{0a}(g-g_{ab})], \quad  \omega_2(k=0)\neq 0
\lab{om2bec}
\ee
Here it should be underlined that
this gapped branch arises definitely due to the external driving. In fact,
without such driving both branches are gapless, e.g. $ \omega_{2}(\Omega_R=0)\approx c_2 k+O(k^3)$, where
$c_2=\sqrt{\rho_{0a}(g-g_{ab})/m}$ is the sound velocity.
In above equations we have used following explicit expressions for the self energies:
\be
\ba
X_1=2g\rho_{0a}+\frac{\Omega_R}{2},  \quad \quad X_2=\frac{\Omega_R}{2},\\
X_5=2 g_{ab}\rho_{0a}-\frac{\Omega_R}{2}, \quad \quad       X_6=-\frac{\Omega_R}{2}
\lab{xlarbec}
\ea
\ee
Therefore,  Gaussian approximation possesses  a nice feature: The same chemical potential, corresponding to the minimum
of the free energy is also relevant both for the spectrum and  HP theorem. However, an extension
of this approach e.g. by taking into account anomalous density, $\sigma\sim <\tilde {\psi}\tilde {\psi}>$,
faces a problem, referred as Hohenberg - Martin dilemma  \ci{hm1965,ourhm}, which declines the existence of a universal
chemical potential in a BEC phase. The solution of this problem has been found in Refs. \ci{yukkl,ouryuk2},
by introducing two chemical potentials. We shall discuss such extension in our forthcoming paper.
Here we note that,
the dispersions \re{om1bec}, \re{om2bec} , being in a complete agreement with
results by Cappelaro et al.   \ci{salas17},  coincide with the results by other theoretical
works e.g. \ci{lelloch13,abad13,aftalion16}
in the Bogoliubov approximation with the assumption  $\rho_{0a}\approx \rho_a=\rho/2$.

In reality, especially at finite temperatures, the Bogoliubov approximation
becomes irrelevant. So, in practical calculations within present approximation one may evaluate
the condensate fraction from the normalization condition
\be
\rho_{0a}=\rho/2-\rho_{1a}
\lab{normbec}
\ee
 which is, in fact, a nonlinear algebraic equation with respect to $\rho_{0a}$
with a fixed $\rho$ and $\rho_{1a}$ given by Eqs. \re{rho1aany},\re{om1bec} - \re{xlarbec}.
It is naturally expected that, the solutions should diminish by increasing the temperature
to reach zero at the critical one.

\section{Critical temperature and its shift}
It is well known that MFT is not enough good to describe critical properties of even
one component Bose gases
near BEC transition \ci{andersen}. One of the reasons of such failure is that, due to the Bogoliubov shift
\re{shift}  formulas for an observable parameter include a factor like $g\rho_0$, which
goes to zero near $T_c$ due to $\rho_0(T=T_c)=0$. As a result, a system of particles
"forgets" about the presence of interparticle interaction and behaves like an ideal gas
with $T_c=\Tcnol $ and $\mu(T=T_c)=0$. Particularly, the present approximation also fails
to predict e.g. the shift of the critical temperature due to interparticle interaction.
However, this is not the case for a one body interaction. Below we derive explicit
expressions for $T_c$ as well as for its shift due to this interaction, which can be realized
e.g. by Rabi coupling.
Here we note that, one may also introduce an effective chemical potential $\mu_{eff}=\mu+\Omega_R/2$,
so that  Hugenholtz - Pines relations could be rewritten in a usual way as $\Sigma_n- \Sigma_{an}=\mu_{eff} $
and $\mu_{eff}(T=T_c-0)=0$.
\subsection{Equation for $T_c$}
It is understood that this coupling modifies an effective
chemical potential of the system. In previous section we have shown that,
to evaluate $\mu$ one has to solve the algebraic equation \re{rho1atbig} or \re{musym} and
 \re{normbec}
for the normal $T>T_c$ or BEC $T<T_c$ phases, respectively. But how about $\mu(T=T_c)$?
Actually, it can be found, for example,  from continuity condition of the self energies as
 $X_1(T=T_c-0)=X_1(T=T_c+0)$ and $X_2(T=T_c-0)=X_2(T=T_c+0)$. So, by using Eqs. \re{Xlartbig}
and \re{xlarbec} we immediately find
\be
\ba
X_1(T=T_c)=X_2(T=T_c)=-\mu(T=T_c)=-\dsfrac{\Omega_R}{2}\\
X_5(T=T_c)=X_6(T=T_c)=-\dsfrac{\Omega_R}{2}
\lab{mutc}
\ea
\ee
and inserting these into \re{dispany}  recognize that there are still two different  branches
of the spectrum:
\be
\omega_1(T=T_c)=\vepsk, \quad \quad
\omega_2(T=T_c)=\vepsk+\Omega_R
\lab{omtc}
\ee
 Now using Eqs. \re{mutc}  and \re{omtc} in \re{rho1aany} we obtain following equations for the critical
temperature $T_c\equiv 1/\beta_c$:
\bea \nonumber
\rho_{1a}(T=T_c)=\rho_a=\rho/2= 
\dsfrac{1}{2V}\dssum_k\left[\dsfrac{1}{e^{\varepsilon_k/\beta_c}-1}+
\frac{1}{z^{-1}e^{\varepsilon_k/\beta_c}-1}
\right]
\lab{rhotc}
\eea
where $z=\exp(-\Omega_R/T_c)$ is the fugacity due to the Rabi coupling, $0\leq z \leq 1$.
As to the $\rho$ in this equation, it is clear that,
 when the Rabi coupling is switched on or off the density
of the whole uniform system remains unchanged: $\rho=\rho(\Omega_R=0)=\rho(\Omega_R\neq 0)$,
being as an fixed input parameter. So, bearing in mind this trivial condition
one can define $\Tcnol$ as the solution to the  following equation:
\be
\rho_a=\rho/2=
\dsfrac{1}{V}\dssum_k\left[\dsfrac{1}{e^{\varepsilon_k/\Tcnol}-1}
\right]
\lab{deftc0}
\ee
which is just the particular case of \re{rhotc} for $\Omega_R=0$. Thus we obtain following
 relation between   $T_c$ and $\Tcnol$
\bea \nonumber
\frac{1}{2}\sum_k\left[\frac{1}{e^{\varepsilon_k/T_c}-1}+\frac{1}{e^{\varepsilon_k/T_c}z^{-1}-1}\right]
-
\sum_k\frac{1}{e^{\varepsilon_k/T_c^0}-1}=0
\lab{eqtctc0sum}
\eea
with $\sum_k\equiv V\int d^3 k/(2\pi)^3$.
This relation can be presented in a rather compact form as
\be
T_c=T_c^0\left[\frac{2\zeta(3/2)}{[\zeta(3/2)+\text{Li}(3/2,z)]}\right]^{2/3}
\lab{tctc0rel}
\ee
by  using following  well known integrals
\be
\ba
 \dsint \dsfrac{ d^3 k}
{
(2\pi)^3 [e^{\varepsilon_k/T}-1]
}
={\tilde c}(mT)^{3/2}\\
\dsint  \dsfrac{ d^3 k}{(2\pi)^3 [z^{-1}e^{\varepsilon_k/T}-1]}=\frac{{\tilde c}(mT)^{3/2}}{\zeta(3/2)}$Li$(3/2,z)
\lab{ints}
\ea
\ee
where ${\tilde c}=\zeta(3/2)/2\sqrt{2}\pi^{3/2}$ ,  $\zeta(3/2)=$Li$(3/2,1)$ and Li$(3/2,z)$ is
the polylogarithm function.
The equation \re{tctc0rel} is a nonlinear algebraic  equation with respect to $T_{c}$ with given input parameters $\Tcnol$ and $\Omega_R$.
In the next subsection we will discuss its exact as well as approximative solutions.
\subsection{The shift vs $\Omega_R$}
From Eq. \re{tctc0rel} one determines the  the shift of the critical temperature due to Rabi coupling as:
\be
\Delta_R=\frac{\Delta T_c}{T_c^0}=\frac{T_c-T_c^0}{T_c^0}=\left[\frac{2\zeta(3/2)}{[\zeta(3/2)+\text{Li}(3/2,z)]}\right]^{2/3}-1
\lab{shiftdfelta}
\ee
In Figs.1 we presented $\Delta_R$ vs $\Omega_R/\Tcnol$.
\begin{figure}[h]
\begin{minipage}[h]{0.48\linewidth}
  \includegraphics[clip,width=\columnwidth]{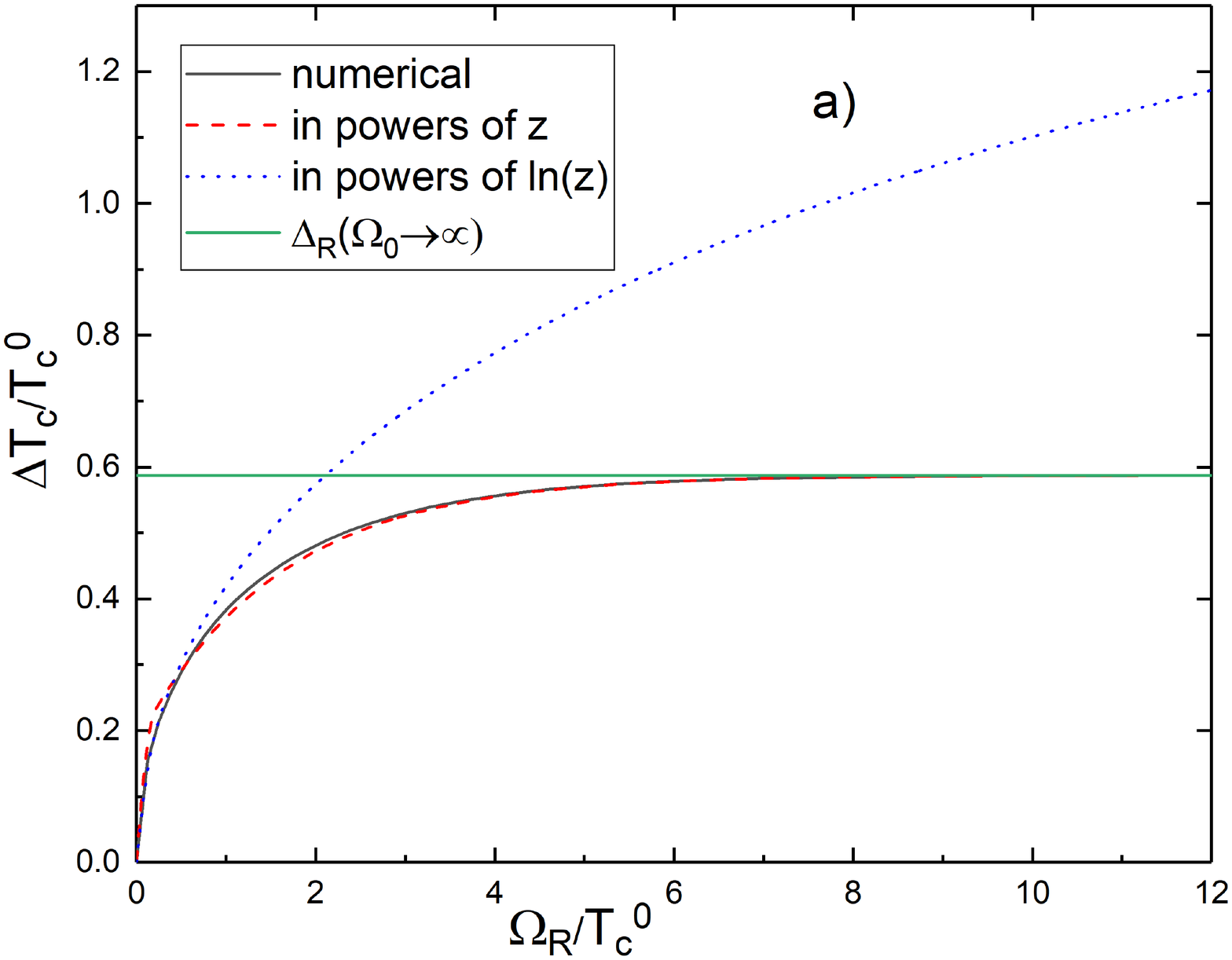}%
\\
\end{minipage}
\hfill
\begin{minipage}[h]{0.48\linewidth}
  \includegraphics[clip,width=\columnwidth]{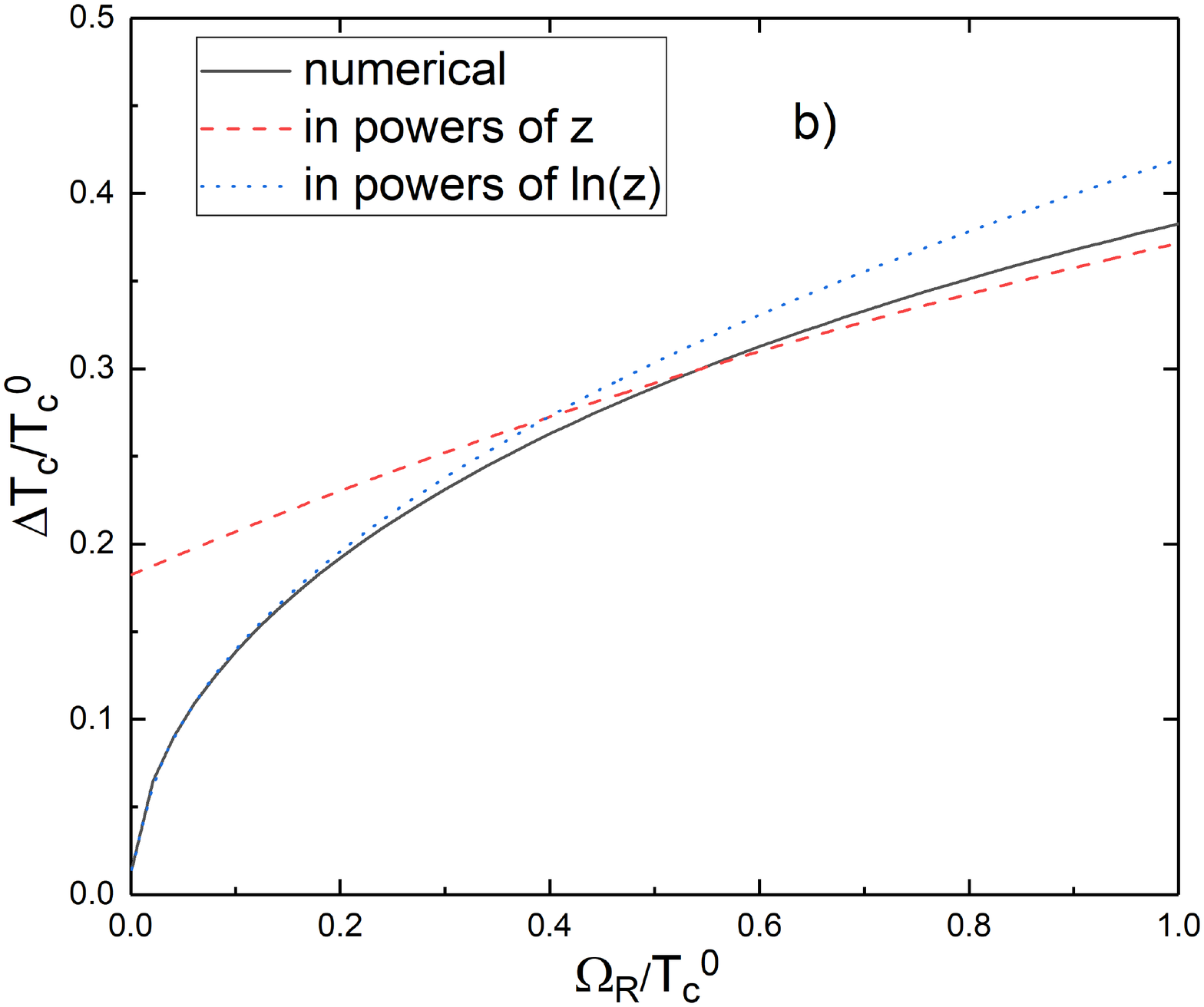}%
\\
\end{minipage}
\hfill

\caption{(a) 
The shift of critical temperature due to Rabi coupling vs the dimensionless parameter $\Omega_R/\Tcnol$. The results
of the exact solution of the Eq. \re{tctc0rel}, expansion in powers of $z=\exp(-\Omega_R/\Tcnol)$,
\re{bigz}
 and
in powers of $\ln(z)$, \re{expsmz} are presented by solid, dashed and dotted curves , respectively.
(b)  The same as in (a) but in a rather larger scale for $\Omega_R/\Tcnol\leq 1$
}
  \label{Fig1}
\end{figure}

 As it is seen from Figs.1 the shift is always positive and tends to a plateau at large $\Omega_R$
for a finite $\Tcnol$. This asymptotic value can be easily evaluated from the Eq. \re{tctc0rel}. In fact, since
$z(\Omega_R\rightarrow \infty)=0$, and Li$(3/2,0)=0$ one obtains  $\Delta_R (\Omega_R\rightarrow \infty)=2^{2/3}-1\approx 0.5874$.
The dashed and dotted curves in Figs. 1 are approximative solutions of Eq. \re{tctc0rel},
while the solid one is the exact numerical solution.

In the literature one can find two kinds of approximations for Li$(3/2,z)$ (see e.g. \ci{wood}):
\be
\ba
\text{Li}(3/2,z)=z+\dsfrac{\sqrt{2}z^2}{4}+\dsfrac{\sqrt{3}z^3}{9}+O(z^4)
\label{expandz}
\ea
\ee
and
\be
\ba
\text{Li}(3/2,z)=-2\sqrt{\pi}\sqrt{\ln(1/ z)}+\zeta(3/2)+\\
\zeta(1/2)\ln (z)+\dsfrac{\zeta(-1/2)}{2}\ln^2 (z)+O(\ln^3 (z))
\label{expandlnz}
\ea
\ee
The former \re{expandz}  is good for small $z<0.3$ ( that is for large $\Omega_R$) and
the latter \re{expandlnz} for rather large $0.3\leq z \leq 1$ (  for small $\Omega_R$). Thus using these
expansions in \re{tctc0rel} and solving the equation iteratively we find following approximations
\be
\ba
\Delta_R (\Omega_R/\Tcnol\leq 1)=\dsfrac{432^{1/3}\sqrt{\pi}2^{2/3}}{18\zeta(3/2)}\sqrt{x}+
\dsfrac{\pi+\zeta(1/2)\zeta(3/2)}{3(\zeta(3/2))^2}x+O(x^{3/2})\approx 
0.45\sqrt{x}-0.033 x
\lab{expsmz}
\ea
\ee
for small  and
\be
\ba
\Delta_R(\Omega_R/\Tcnol >>  1)\approx -\dsfrac{2^{5/2}\exp(-\frac{x}{2^{2/3}})        }
{3\zeta(3/2)}+
2^{2/3}-1\approx
0.5874-0.4\exp({-{0.63 x}})
\lab{bigz}
\ea
\ee
 large values of the dimensionless parameter $x\equiv \Omega_R/\Tcnol$. This is illustrated in Fig.1b,
where dashed and dotted curves are appropriate for the Eqs. \re{bigz} and \re{expsmz}
respectively.

Note that, existing experiments are performed with $\Omega_R\sim 1000 Hz\approx 0.48\cdot 10^{-7}$K at temperatures
$T\sim 2 \cdot 10^{-7}$ K , which corresponds to rather small values of ${\Omega_R}/{T_c^0}\sim 0.25$. From Fig.1b
we conclude that, in reality one may use the approximation given by Eq. \re{expsmz} which gives
about $20\%$ near this region  for the shift of the critical temperature due to Rabi coupling.

\section{Conclusion}
In present work we have studied Rabi coupled binary Bose system within the Gaussian (bilinear)
approximation in the framework of the mean field theory. We have shown that for this system
at finite temperatures BEC can exist  only for a symmetric case
(  $g_a=g_b$) with an equal population, otherwise only a crossover transition may
take place. For the first time we have derived Hugenholtz - Pines relations
and an equation for the shift of the critical temperature in the presence
of one body interaction.  Although our extended HP relations: $\Sigma_n-\Sigma_{an}=\mu+\Omega_R/2$,
and $\Sigma_{n}^{ab}=\Sigma_{an}^{ab}$ are proved within the field theoretical Gaussian approximation,
we expect that they will remain true beyond this approximation also and can be proved
more accurately, say in the spirit of works \ci{watabe21,bogpines}

As to the critical temperature $T_c$ of a BEC transition in Rabi coupled two component system, it has
never been studied before. So,  we obtained rather simple equation to determine $T_c$ as well as its shift due to
Rabi coupling. The solutions of this equation may be apparently  presented analytically
for small (or large) values of the input parameter $\Omega_R/\Tcnol$.   The shift, as we predicted,
is positive and should be about $\sim 20 \%$ in existing experimental measurements with
the moderate value of parameter $\Omega_R/\Tcnol\sim 0.2$. Principally, one may assume possibility
of  making measurements with rather large values of $\Omega_R/\Tcnol$ also. For such experiments, we predict that
the shift should go to its asymptotic value $\Delta T_c/\Tcnol \leq 2^{2/3}-1\approx 0.58$
, displaying a plateau. We hope that, this prediction can be verified in future experiments
performed at finite temperatures.
On the other hand it would be also   interesting to
estimate the shift by using quite other approximations with a different
perturbation scheme,  for example, used  in Refs. \ci{baym99,kastening04}.

Moreover, we have proved  that, physical observables , particularly  $\Delta T_c/\Tcnol$
do not depend on the sign of one body interaction i.e. on the phase of Rabi coupling $\omega_R$.
Possible effect of changing the sign of this interaction could be reflected only
in the relative phase of the two condensates  $\xi$ under the condition $\xi\omega_R=-1$.

Note that, other problems concerning these systems, such as the temperature dependence of
thermodynamic parameters, as well as stability conditions \ci{salas17} are left beyond the scope
of the present work. We shall study these problems in our forthcoming paper, where we use
a more accurate approximation than the Gaussian one
and make an attempt to
extend the equation \re{hp1} for a general asymmetric binary Bose system with a crossover \ci{ouraniz1}.

\section*{Acknowledgments}
We are indebted to S. Watabe and  V. Yukalov  for useful discussions.

\clearpage
\appendix
\centerline{\bf APPENDIX}
\section{ Relation between the phases and phase invariance}
\numberwithin{equation}{section}
\setcounter{equation}{0}
In present Appendix we show that, the stability condition with respect to
the relative phase $\xi=\exp(i\theta)$
 leads to  the relation $\xi\omega_R=-1$.
In fact,  the relative  phase angle $\theta$, should  correspond to the minimum of $\Omega_0$:
\be \label{equlib2}
\frac{\partial\Omega_0}{\partial\theta}=0, \quad \frac{\partial^2\Omega_0}{\partial\theta^2}>0
\ee
In general, $\Omega_0$ is given by
\be
\ba
\Omega_0=V[-\mu\rho_{0a}-\mu\rho_{0b}+\dsfrac{g_a\rho_{0a}^2}{2}+\dsfrac{g_b\rho_{0b}^2}{2}+
g_{ab}\rho_{0a}\rho_{0b}+\dsfrac{\Omega_R\sqrt{\rho_{0a}\rho_{0b}}\xi(\omega_R+\omega_R^\star)}{2}]
\lab{om0app}
\ea
\ee
Besides, in stable equilibrium, the variational parameters $\rho_{0a}$ and $\rho_{0b}$ should
satisfy the saddle-point equations \ci{salas17,hague,andersen}
\bea \label{equlib} \nonumber
\frac{\partial\Omega_0}{\partial\rho_{0a}}=0, \quad \frac{\partial\Omega_0}{\partial\rho_{0b}}=0 \quad \\ \left(\frac{\partial^2\Omega_0}{\partial\rho_{0a}^2}\right)\left(\frac{\partial^2\Omega_0}{\partial\rho_{0b}^2}\right)-\left(\frac{\partial^2\Omega_0}{\partial\rho_{0a}\partial\rho_{0b}}\right)>0
\eea
Now bearing in mind $\omega_R=e^{-i\theta_R}$, and using Eqs. \re{equlib2},  \re{om0app},  we obtain
\be
\begin{split}
\frac{\partial\Omega_0}{\partial\theta}=-V\Omega_R\sqrt{\rho_{0a}\rho_{0b}}\sin(\theta-\theta_R)=0 \\
\frac{\partial^2\Omega_0}{\partial\theta^2}=-V\Omega_R\sqrt{\rho_{0a}\rho_{0b}}\cos(\theta-\theta_R)>0
\end{split}
\ee
which lead to the well known \ci{Pitbook14,lelloch13,abad13}   result:
$\theta-\theta_R=\pi(2n+1)$ i.e. $\omega_R\xi=-1$. This physically means that, when the system goes into
its equilibrium state, it will choose the relative phase  between the two condensates by itself,
preferring the case with $\xi=-\omega_R$, where $\omega_R$ is in fact the sign of the one body interaction.
For example, if the latter is negative, $\omega_R=-1$ , then the condensates will coexist
with  the same phase $\xi=+1$, and vise versa.
Moreover, particularly , in both cases, with $\xi=\pm 1$ there always exist in phase and out of phase
excitations. The former correspond to the density branch $\omega_d$, while the latter
to the spin branch $\omega_s$ , as it has been  clarified in Refs. \ci{ouryuk2,kim20}

Below we show that physical observables do not depend on the relative phase $\xi$.
Actually,
in the symmetric case for the self energies we have
\be \label{Xn}
\begin{split}
&X_1=X_3=-\mu+3\rho_{0a}g+\rho_{0a}g_{ab}=\Sigma_n+\Sigma_{an}-\mu
\\
&X_2=X_4=-\mu+\rho_{0a}g+\rho_{0a}g_{ab}=\Sigma_n-\Sigma_{an}-\mu
\\
&X_5=2\xi g_{ab}\rho_{0a}+\frac{\Omega_R\omega_R}{2}, \quad X_6=\frac{\Omega_R\omega_R}{2}
\end{split}
\ee
and for the  dispersions:
\be
\ba
\omega_1=\sqrt{(\varepsilon_k+X_2+X_6)(\varepsilon_k+X_1+X_5)}  \\
\omega_2=
\sqrt{(\varepsilon_k+X_2-X_6)(\varepsilon_k+X_1-X_5)}
\ea
\label{dispany}
\ee
Now from Eqs. \re{rho1aany},  \re{Xn} and \re{dispany}
it is seen that e.g. $\rho_{1a}$ is phase invariant, that is  $\rho_{1a}(\xi,\omega_{R})=\rho_{1a}(-\xi,-\omega_{R})$,
since the transformation  $(\xi,\omega_{R})\leftrightarrow(-\xi,-\omega_{R})$
is equivalent to the following replacements:
 $X_5\leftrightarrow -X_5$,  $X_6\leftrightarrow -X_6$,
$\omega_1\leftrightarrow \omega_2$.
Note that, this statement holds true both for the condensed as well as normal states.
Phase invariance of other observables can be proven in the same way.
\bibliography{sn-bibliography}


\end{document}